\def\year{2019}\relax
\documentclass[letterpaper]{article} 
\usepackage{aaai19}  
\usepackage{times}  
\usepackage{helvet}  
\usepackage{courier}  
\usepackage{url}  
\usepackage{graphicx}  
\usepackage{multirow}
\usepackage{graphicx}
\usepackage{enumitem}
\usepackage{array}
\usepackage{amssymb}
\usepackage{mathtools}
\usepackage{xcolor}
\usepackage{pgfplots}
\pgfplotsset{compat=newest}
\usepackage{filecontents}
\usepackage{tikz}
\usepackage{subcaption}

\newcolumntype{L}[1]{>{\raggedright\let\newline\\\arraybackslash\hspace{0pt}}m{#1}}
\newcolumntype{C}[1]{>{\centering\let\newline\\\arraybackslash\hspace{0pt}}m{#1}}
\newcolumntype{R}[1]{>{\raggedleft\let\newline\\\arraybackslash\hspace{0pt}}m{#1}}

\newcommand{\an}[1]{$^{{\textrm{\tiny #1}}}$}
\newcommand{\citet}[1]{\citeauthor{#1} (\citeyear{#1})}

\begin{document}
\title{Multi-Perspective Relevance Matching with \\ Hierarchical ConvNets for Social Media Search}
\author{Jinfeng Rao,\an{1}\thanks{Work done at the University of Maryland, College Park.} Wei Yang,\an{2} Yuhao Zhang,\an{3} Ferhan Ture,\an{4} and Jimmy Lin\an{2}\\[0.1cm]
$^{1}$ Facebook Conversational AI \\
$^{2}$ David R. Cheriton School of Computer Science, University of Waterloo \\
$^{3}$ Department of Computer Science, Stanford University \\
$^{4}$ Comcast Applied AI Research \\[0.1cm]
raojinfeng@fb.com, \{w85yang,jimmylin\}@uwaterloo.ca, yuhao.zhang@stanford.edu, ferhan\_ture@comcast.com
}

\maketitle

\begin{abstract}
Despite substantial interest in applications of neural networks to
information retrieval, neural ranking models have mostly been applied to
``standard'' {\it ad hoc} retrieval tasks over web pages and newswire
articles. This paper proposes MP-HCNN ({\bf M}ulti-{\bf
  P}erspective {\bf H}ierarchical {\bf C}onvolutional {\bf N}eural
{\bf N}etwork), a novel neural ranking model specifically designed for
ranking short social media posts. We identify document length,
informal language, and heterogeneous relevance signals as features
that distinguish documents in our domain, and present a model
specifically designed with these characteristics in mind. Our model
uses hierarchical convolutional layers to learn latent semantic soft-match
relevance signals at the character, word, and phrase levels. A
pooling-based similarity measurement layer integrates evidence
from multiple types of matches between the query, the social media
post, as well as URLs contained in the post. Extensive experiments using
Twitter data from the TREC Microblog Tracks 2011--2014 show that our
model significantly outperforms prior feature-based as well as
existing neural ranking models. To our best knowledge, this paper
presents the first substantial work tackling search over social media
posts using neural ranking models. Our code and data are publicly available.\footnote{https://github.com/jinfengr/neural-tweet-search}
\end{abstract}

\section{Introduction}

In recent years, techniques based on neural networks offer exciting
opportunities for the information retrieval (IR) community. For example,
distributed word representations such as
word2vec~\cite{mikolov2013distributed}~provide a promising basis to overcome
the longstanding vocabulary mismatch problem in ranking~\cite{ganguly2015word},
which refers to the phenomenon where queries and documents describe the same concept with different words.
Nevertheless, there are still fundamental challenges to be solved. 
Guo et al.~\shortcite{guo2016deep} pointed out that
\emph{relevance matching}, which is the core problem in IR,
has different characteristics from the \emph{semantic matching} problem 
that many NLP models are designed for, which is essentially to model 
how semantically close two pieces of texts are, such as paraphrase detection~\cite{socher2011dynamic} and answer sentence selection~\cite{rao2016noise}. 
In particular, exact match
signals still play a critical role in ranking, more than the role
of term matching in, for example, paraphrase detection.
Furthermore, in document ranking there is an asymmetry between
queries and documents in terms of length and the richness of signals that can be
extracted; thus, symmetric models such as Siamese architectures may
not be entirely appropriate.
Nevertheless, significant progress has been made, and many neural
ranking models have been recently proposed~\cite{shen2014learning,huang2013learning,pang2016text,xiong2017end},
which have been shown to be effective for {\it ad hoc} retrieval.

Despite much progress, it remains unclear how neural ranking models
designed for ``traditional'' {\it ad hoc} retrieval tasks perform on
searching social media posts such as tweets on Twitter. We can
identify several important differences:

\begin{itemize}[leftmargin=*]

\item {\bf Document length}. Social media posts are much shorter than web or
newswire articles. For example, tweets are limited to 280 characters.
Thus, {\it ad hoc} retrieval in this domain contains elements of semantic matching because
queries and posts are much closer in length.
In particular, neural models that rely on
paragraph-level interactions and global matching mechanisms~\cite{mitra2017learning} are unlikely to be effective.

\item {\bf Informality}. 
Idiosyncratic conventions (e.g., hashtags), abbreviations (``Happy Birthday'' as ``HBD''), 
typos, intentional misspellings, and emojis are prevalent in social media posts.
An effective ranking model should account for 
such language variations and term mismatches due to the informality of posts.

\item {\bf Heterogeneous relevance signals}. The nature of social media platforms drives
users to be actively engaged in real-world news and events;
users frequently take advantage of URLs or hashtags to increase exposure to their posts.
Such heterogeneous signals are not well exploited by existing models,
which can potentially boost ranking effectiveness when modeled together with textual content.

\end{itemize}

\noindent 
We present a novel neural ranking model for {\it ad hoc}
retrieval over short social media posts that is specifically designed
with the above characteristics in mind. Our model,
MP-HCNN ({\bf M}ulti-{\bf P}erspective {\bf H}ierarchical 
{\bf C}onvolutional {\bf N}eural {\bf N}etwork), aims to model the
relevance of a social media post to a query in a multi-perspective
manner, and has three key features:

\begin{enumerate}[leftmargin=*]

\item To cope with the informality of social media and to
  support more robust matching, we apply word-level as well as
  character-level modeling, with URL-specific matching. This
  allows us to exploit noisy relevance signals at different
  granularities.

\item Our model consists of hierarchical convolutional layers to capture 
multi-level latent soft-match signals between query and post contents,
starting from character-level and word-level to phrase-level, and finally to sentence-level.

\item Matching of learned representations between query and posts as
  well as URLs is accomplished with a similarity
  measurement layer where term importance weights are injected at
  each convolutional layer as priors.

\end{enumerate}

\noindent Finally, all relevance signals are integrated using a
fully-connected layer to yield the final relevance
ranking. Optionally, neural matching scores can be integrated with
lexical matching via linear interpolation to further improve ranking.

\smallskip \noindent {\bf Contributions.} We view our contributions as follows:

\begin{itemize}[leftmargin=*]

\item We highlight three important characteristics of social media
  posts that make {\it ad hoc} retrieval over such collections
  different from searching web pages and newswire articles. Starting
  from these insights, we developed MP-HCNN, a novel neural
  ranking model specifically designed to address these
  characteristics. To our best knowledge, ours is the first neural ranking 
  model developed specifically for {\it ad hoc} retrieval over social media posts.

\item We evaluate the effectiveness of our MP-HCNN model on
  four Twitter benchmark collections from the TREC Microblog Tracks
  2011--2014. Our model is compared to learning-to-rank approaches as
  well as many recent state-of-the-art neural ranking models that are
  designed for web search and ``traditional'' {\it ad hoc} retrieval. Extensive
  experiments show that our model significantly improves the state of the art over
  previous approaches. Ablation studies further confirm that
  these improvements come from specific components of our
  model designed to tackle characteristics of social media posts
  identified above.

\end{itemize}	

\section{Related Work}

Deep learning has achieved great success in many natural language processing and information
retrieval applications~\cite{sutskever2014sequence,yin2015abcnn,he2016pairwise,rao2017talking}.
Early attempts at neural IR
mainly focus on representation-based modeling between query and document,
such as DSSM~\cite{huang2013learning}, C-DSSM~\cite{shen2014learning}, 
and SM-CNN~\cite{severyn2015learning}. DSSM is an early
NN architecture for web search that maps word sequences to character-level trigrams 
using a word hashing layer, and then feeds the dense hashed features to a multi-layer perceptron (MLP) 
for similarity learning. C-DSSM extends this idea by replacing the MLP
in DSSM with a convolutional-based CNN to capture local contextual signals from neighboring character trigrams.

More recently, interaction-based approaches~\cite{guo2016deep,xiong2017end,mitra2017learning,dai2018convolutional}
have demonstrated increased effectiveness in many ranking tasks.
They operate on the similarity matrix of word pairs from 
query and document, which is usually computed through word embeddings such as word2vec~\cite{mikolov2013distributed}.
The DRMM model~\cite{guo2016deep}
introduces a pyramid pooling technique to convert the similarity matrix to 
histogram representations, on top of which a term gating
network aggregates weighted matching signals from different query terms. 
Inspired by DRMM, \citet{xiong2017end} propose K-NRM, which
introduces a differentiable kernel-based pooling technique to capture matching
signals at different strength levels. \citet{dai2018convolutional} extends
this idea to model soft-match signals for $n$-grams with an additional convolutional layer.  
The DUET model~\cite{mitra2017learning} combines representation-based and interaction-based
techniques with a global component for semantic matches and a local component for exact matches.

Our model differs from previous work in a number of ways:
(1) we motivate the need for character-level modeling of noisy 
texts and URLs in social media and provide a tailored design for
this purpose; 
(2) we organize convolutional layers in a hierarchical manner to better model the semantics of words and phrases, 
and found it to be more effective than previous architectures;
(3) we propose a
parameter-free similarity measurement mechanism coupled with external weights 
to capture multiple levels of term matching signals, which provides our model better interpretability. 
Detailed ablation experiments verify
the contributions of various components in our architecture.

\section{Multi-Perspective Model}	

\begin{figure*}
\centering
\includegraphics[width=0.98\linewidth]{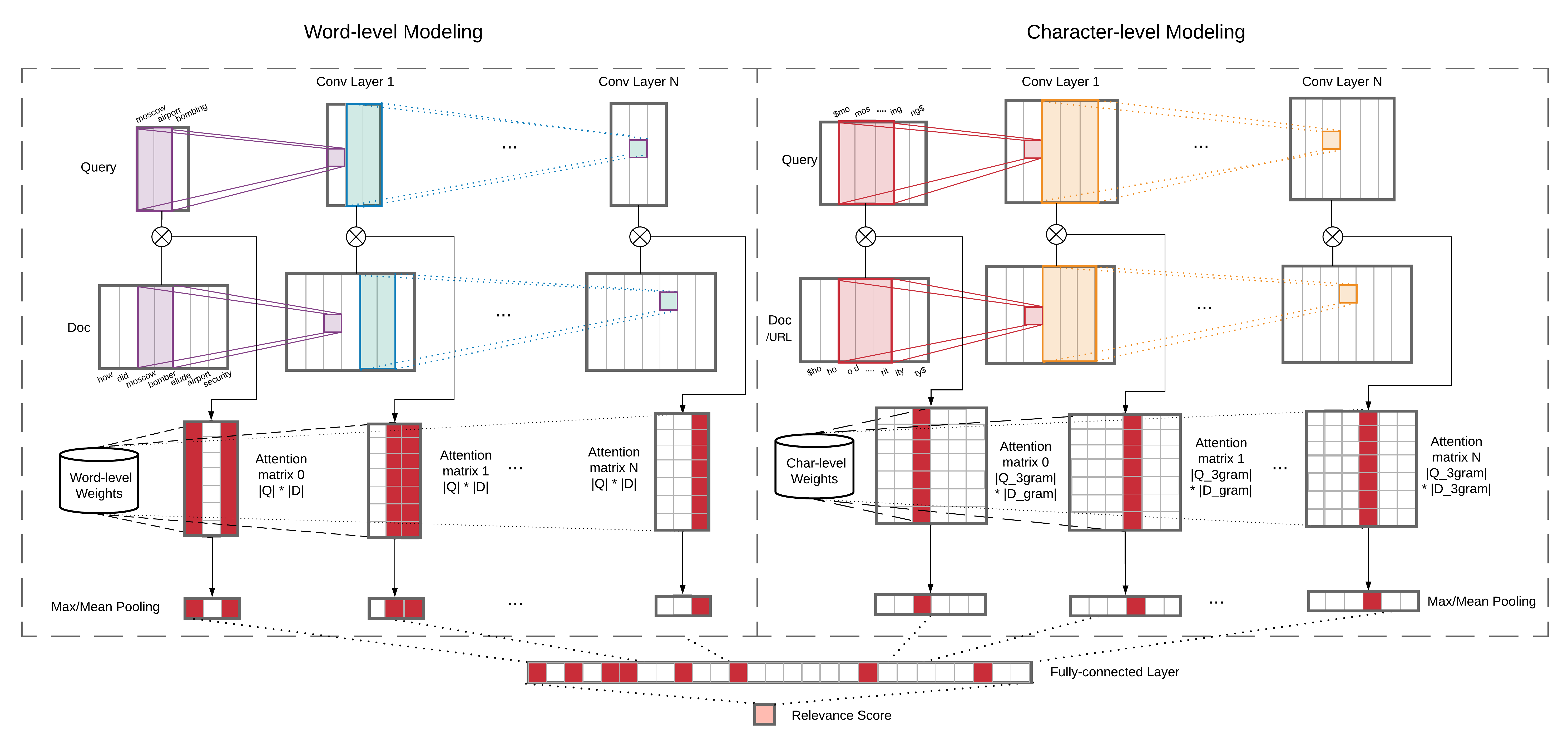}
\caption{Overview of our Multi-Perspective Hierarchical Convolutional Neural
Network (MP-HCNN), which consists of two parallel components for
word-level and character-level modeling between queries, social media
posts, and URLs.  The two parallel components share the same
architecture (with different parameters), which comprises hierarchical
convolutional layers for representation learning and a semantic
similarity layer for multi-level matching.  Finally, all relevance
signals are integrated using a fully-connected layer to produce the
final relevance score. }  \label{fig:model}
\end{figure*}

The core contribution of this paper is a novel neural ranking model
specifically designed for {\it ad hoc} retrieval over short social
media posts.  
As discussed in the introduction, our model,
MP-HCNN (Multi-Perspective Hierarchical Convolutional Neural
Network), has three key features: First, we apply word-level as well
as character-level modeling on query, posts, and URLs to cope with the
informality of social media posts.
Second, we exploit stacked convolutional layers to
learn soft-match relevance at multiple granularities.  Finally, we learn matches between
the learned representations via pooling with injected external weights. Our overall model architecture
is shown in Figure~\ref{fig:model}, and each of the above key features
are described in detail below.
	
\subsection{Multi-Perspective Input Modeling}
\label{sec:input-modeling}

A standard starting point for neural text processing is to take advantage of word embeddings,
e.g., word2vec~\cite{mikolov2013distributed}, to encode each word. However, in the social media domain, 
informal post contents contain many out of vocabulary (OOV) words
which can't be found in pre-trained word embeddings. The embeddings of OOV words
are randomly initialized by default. In fact, we observe that
about 50\%--60\% of words are OOV in the TREC Microblog datasets
(details in Table~\ref{statistics}). This greatly complicates the matching process if we simply rely on
word-level semantics, thus motivating the need for character-level input modeling to 
cope with noisy texts.

To better understand the source of OOV words,
we randomly selected 500 OOV words from the vocabulary and 
provide a few examples below of the major sources of OOV occurrences in the social media domain:

\begin{enumerate}
	\item \textbf{Compounds} (42.4\%): chrome-os, actor-director
	\item \textbf{Non-English words} (29.2\%): emociones (Spanish, emotions), desgostosa (Portuguese, disgusted)
	\item \textbf{Typos} (17.1\%): begngen (beggen), yawnn (yawn), tansport (transport), afternoo (afternoon), foreverrrr (forever)
	\item \textbf{Abbreviations} (5.6\%): EASP (European Association of Social Psychology), b-day (birthday)
	\item \textbf{Domain-specific words} (5.7\%): utf-8, vlookup
\end{enumerate}

\noindent As we can see above, compounds, non-English words, and typos are the three biggest sources of OOV words.
Character-level modeling is beneficial for both the compounds and typos cases.

In addition, social media posts often contain many heterogeneous features that can contain 
fruitful relevance signals, such as mentions, hashtags, and external URL links. An analysis of
the TREC Microblog Track 2011--2014 datasets shows that around 50\% of tweets contain one
or more URLs. More detailed statistics can be found in Table~\ref{statistics}. In fact, by taking
a closer look at real data, we see that many URL links can be fuzzy matched to query terms.
We provide one example in Table~\ref{example}. For those posts without URLs, we add a placeholder 
symbol ``$<$URL$>$''. Note that while the HTML pages referenced by the URLs are another
obvious source of relevance signals, we do not consider them in our model because many of
those URLs are no longer accessible, and noisy HTML documents require additional preprocessing,
which is beyond the scope of this paper.

\begin{table}
	\centering
	\begin{tabular}{ll}
	\hline
	Topic & 1: BBC world service cuts \\
	\hline
	Tweet & BBC slashes online budget by 25\% will \\ 
	& cut 360 employees and 200 websites \#bbcnews. \\
	\hline
	URL & http://bbc-world-service-to-cut-staff.html \\

	\hline
	\end{tabular}
	\caption{Example query-post pair retrieved by topic 1 from the TREC Microblog 2011 dataset.}
	\label{example}
\end{table}
    
To tackle the language variation issues discussed above and to exploit URL information, 
we consider multiple inputs for relevance modeling: (1) query and post at word-level; (2)
query and post at character-level; (3) query and URL at character-level. 
For character-level modeling, we segment the
query and post contents as well as the URL link to a sequence of character trigrams 
(e.g., ``hello" to \{\#he, hel, ell, llo, lo\#\}), 
which has been shown to yield good effectiveness in capturing morphological
variations~\cite{huang2013learning}.
We adopt the same architecture to capture word-level semantic and character-level
matching signals, discussed next.
	
\subsection{Hierarchical Representation Learning}
\label{sec:text-matching}

Given a query $q$ and a document $d$,
the textual matching component aims to learn a relevance score
$f(q, d)$ using the query terms $\{w_1^q, w_2^q, ..., w_n^q\}$ and document terms $\{w_1^d, w_2^d, ..., w_m^d\}$,
where $n$ and $m$ are the number of terms in $q$ and $d$, respectively.
To be clear, ``document'' can either refer to a social media post or an URL, and ``term''
refers to either words or character trigrams.
One important novel aspect of our model is relevance modeling from multiple perspectives, and 
our architecture exhibits symmetry in the word- and character-level modeling (see Figure~\ref{fig:model}).
Thus, for expository convenience, we use ``document'' and ``term'' in the generic sense above.
We first employ an embedding layer to
convert each term into a $L$-dimensional vector representation, generating a matrix representation for the query
$\textbf{Q}$ and document $\textbf{D}$, where $\textbf{Q} \in \mathbb{R}^{n \times L}$ and $\textbf{D} \in \mathbb{R}^{m \times L}$. In the following, we introduce our representation learning method with hierarchical convolutional neural networks. 

A convolutional layer applies convolutional filters to the text, which is represented by 
an embedding matrix $\textbf{M}$ ($\textbf{Q}$ or $\textbf{D}$).
Each filter is moved through the input embedding incrementally as a sliding window (with window size $k$) 
to capture the compositional representation of $k$ neighboring terms. Assuming a convolution layer has $F$ filters, then this CNN layer produces output matrix $\textbf{M}^o \in \mathbb{R}^{\|M\| \times F}$ with $O(F \times k \times L)$ parameters.

We then stack multiple convolutional layers in a hierarchical manner to 
obtain higher-level $k$-gram representations.
For notational simplicity, we drop the superscript $o$ from all output matrices and add
a superscript $h$ to denote the output of the $h$-th convolutional layer. 
Stacking $N$ CNN layers therefore corresponds to obtaining the output matrix of the $h$-th layer $\textbf{M}^{h} \in \mathbb{R}^{\|M\| \times F^h}$ via:
$$\textbf{M}^{h} = \textrm{CNN}^h(\textbf{M}^{h-1}), h = 1, \ldots, N ,$$
where $\textbf{M}^{h-1}$ is the output matrix of the $(h-1)$-th convolutional layer.
Note that $\textbf{M}^0 = \textbf{M}$ denotes the input
matrix ($\textbf{Q}$ or $\textbf{D}$) obtained directly from the word embedding layer, and the parameters of 
each CNN layer are shared by the query and document inputs.

Intuitively, consecutive convolutional layers allow us to obtain higher-level abstractions
of the text, starting from character-level or word-level to phrase-level and eventually to sentence-level.
A single CNN layer is able to capture the $k$-gram semantics from the input embeddings,
and two CNN layers together allow us to expand the context window to up to $2k-1$ terms.
Generally speaking, the deeper the convolutional layers, 
the wider the context considered for relevance matching.
Empirically, we found that a filter size $k=2$ for word-level inputs and $k=4$ for character-level inputs work well.
The number of convolutional layers $N$ was set to 4. 
This setting is reasonable as it enables us to gradually learn the representations of 
word-level and character-level $n$-grams of up to $O(N \times k)$ length. 
Since most queries and documents
in the social media domain are either shorter or not much longer than this length, we can regard the output
from the last CNN layer as an approximation of sentence representations. 

An alternative to our deep hierarchical design is a \textit{wide} architecture, which 
reduces the depth but expands the width of the network
by concatenating multiple convolutional
layers with different filter sizes $k$ in parallel to learn variable-sized phrase representations.
However, such a design will require quadratically more parameters and be more difficult to learn than our approach.
More specifically, our deep model comprises $O(N\times F \times kL)$ parameters with
$N$ CNN layers, while a wide architecture with the same representation window 
will need $O(F \times (kL + 2kL + ... + NkL)) = O (N^2 \times F \times kL) $ parameters. 
The reduced parameters in our approach mainly come from representation reuse at each CNN layer,
which also generalizes the learning process by sharing representations between successive layers. 

\subsection{Similarity Measurement and Weighting}
\label{sec:sim-and-weighting}

To measure the similarity between the query and the document, we match the query with the document at each convolutional layer by taking the
dot product between the query representation matrix $\textbf{M}_q$
and the document representation matrix $\textbf{M}_d$:
\begin{equation*} \label{eq2}
\begin{aligned}
\textbf{S} &= \textbf{M}_q {\textbf{M}_d}^T, \textbf{S} \in \mathbb{R}^{n \times m}, \\
\tilde{\textbf{S}}_{i,j} &= \textit{softmax}(\textbf{S}_{i,j}) = \frac{e^{\textbf{S}_{i, j}}}{\sum_{k=1}^m {e^{\textbf{S}_{i, k}}}}
\end{aligned}
\end{equation*}

\noindent where $\textbf{S}_{i,j}$ can be considered the similarity score by
matching the query phrase vector $\textbf{M}_q[i]$ with the document phrase vector $\textbf{M}_d[j]$.
Since the query and document share the same convolutional layers, similar phrases
will be placed closer together in a high-dimensional embedding space and their product will
produce larger scores. The similarity matrix \textbf{S} is further normalized to $\tilde{\textbf{S}}$ with range $[0, 1]$
through a document-side $\textit{softmax}$ function. 

We then apply \textit{max} and \textit{mean} pooling to the similarity matrix to obtain discriminative feature vectors:
\begin{equation*} \label{eq4}
\begin{aligned}
\textit{Max}(\tilde{\textbf{S}}) & = [max(\tilde{\mathbf{S}}_{1, :}), ..., max(\tilde{\mathbf{S}}_{n,:})],  \textit{Max}(\tilde{\textbf{S}}) \in \mathbb{R}^{n};
\\
\textit{Mean}(\tilde{\textbf{S}}) & = [mean(\tilde{\mathbf{S}}_{1, :}), ..., mean(\tilde{\mathbf{S}}_{n, :})], \textit{Mean}(\tilde{\textbf{S}}) \in \mathbb{R}^{n}; 
\end{aligned}
\end{equation*}

\noindent Each score generated from pooling can be viewed as one piece of matching evidence 
for a specific query term or phrase to the document, and its value denotes the importance of the relevance signal. 
 
To measure the relative importance of different
query terms and phrases, we inject external weights as prior information by multiplying
the score after pooling with the weight of that specific query term or phrase.
These are provided as feature inputs to the subsequent learning-to-rank layer, denoted by $\Phi$:
\begin{equation*} 
\begin{aligned}
\Phi = \{ weights(q) \odot \textit{Max}(&\tilde{\textbf{S}}), weights(q) \odot \textit{Mean}(\tilde{\textbf{S}})\} ,
\end{aligned}
\label{eq5}
\end{equation*}

\noindent where $\odot$ is an element-wise product between the weights of 
query terms or phrases with the pooling scores and
$weights(q)[i]$ denotes the weight of the $i$-th term or phrase in the query.
We choose inverse document frequency (IDF)
as our weighting measure; a higher IDF weight implies rarer occurrence in the collection and thus larger
discriminative power. This weighting method also reduces the impact of high matching scores 
from common words like stopwords. 

Our similarity measurement layer has two important properties. First, all the layers,
including matching, softmax, pooling, and weights, have no learnable parameters. Second, 
the parameter-free nature enables our model to be more interpretable and 
to be more robust from overfitting. By matching query phrases with document phrases in a joint manner,
we can easily track which matches contribute more to the final prediction. 
This greatly increases the interpretability of our model, an important benefit as this issue has become a prevalent concern
given the complexity of neural models for IR and NLP applications~\cite{li2015visualizing}. 

\subsection{Evidence Integration}

Given similarity features learned from word-level $\Phi^w$ and character-level $\Phi^c$,
we employ a multi-layer perceptron (MLP) with a ReLU activation in between as our learning-to-rank module:
\begin{equation*}
\begin{split}
o = \textit{softmax} \left( \textbf{MLP}(\Phi^w \sqcup \Phi^c) \right)
\end{split}
\end{equation*}
\noindent where $\sqcup$ is a concatenation operation and the $\textit{softmax}$ function normalizes 
the final prediction to a similarity vector $o$ with its values between 0 and 1.
The training goal is to minimize
the negative log likelihood loss $L$ summed over all samples $(o_i, y_i)$: $L = - \sum_{(o_i, y_i)} \log o_i[y_i], $ 
where $y_i$ is the annotation label of sample $i$.

\subsection{Interpolation with Language Model}
\label{sec:interpolation}

Various studies have shown that neural ranking models are good at capturing \textit{soft}-match signals~\cite{guo2016deep,xiong2017end}.
However, are exact match signals still needed for neural methods?
We examine this hypothesis by adopting a commonly-used linear interpolation
method to combine the match scores of NN-based models with language model scores between
a (query, doc) pair:
\begin{equation} \label{eq6}
\begin{aligned}
\textrm{Score}(q, d) = \lambda \cdot \textrm{NN}(q, d) + (1 - \lambda) \cdot \textrm{LM}(q, d).
\end{aligned}
\end{equation}

\noindent We use query-likelihood (QL)~\cite{ponte1998language} as the language model score here.
The interpolation technique is applied to our multi-perspective model as well as other neural
models we use as baselines in this paper. 
We report both effectiveness with and without interpolation in the experimental section.

\section{Experimental Setup}

\noindent \textbf{Dataset.} To evaluate our proposed model for social media search, 
we choose four Twitter test collections from
the TREC Microblog Tracks in 2011, 2012, 2013, and 2014.
Each dataset contains about 50 queries. 
Following standard experimental procedures~\cite{ounis2011overview}, we evaluate our models in
a reranking task, using as input the top 1000 retrieved documents
(tweets) from a bag-of-words query likelihood (QL) model using
the TREC Microblog Track API.\footnote{https://github.com/lintool/twitter-tools}
Note that the API returns less than 1000 tweets for some queries.
The statistics of the four datasets are shown in Table~\ref{statistics}. 
Since most URLs in the tweets are shortened, for example,
given http://zdxabf we recover the original URL from redirection 
for character-level modeling.  

We use the Stanford Tokenizer tool\footnote{https://nlp.stanford.edu/software/tokenizer.shtml} to segment
the retrieved tweets into token sequences to serve as model input.
Non-ASCII characters are removed.
We run four sets of experiments, where each of the four datasets is used for
evaluation, with the other three used for training (e.g., train on TREC 2011--2013, test on TREC 2014).
In each experiment, we sample 15\% of the training queries as the validation set. 
Following the official track guidelines~\cite{ounis2011overview}, 
we adopt mean average precision (MAP) and precision at 30 (P@30) as our
evaluation metrics. 
The relevance judgments are made on a three-point scale 
(``not relevant'', ``relevant'', ``highly relevant'') and we treat both higher grades as relevant,
per Ounis et al.~\shortcite{ounis2011overview}. 

\begin{table}
	\begin{tabular}{lrrrr}
		\hline
		{\bf Test Set} & {\bf 2011}  & {\bf 2012}  & {\bf 2013}  & {\bf 2014} \\
		\hline \hline
		\# query topics   & 49    & 60    & 60    & 55    \\ 
		\# query-doc pairs & 39,780 & 49,879 & 46,192 & 41,579 \\ 
		\# relevant docs   & 1,940  & 4,298  & 3,405  & 6,812  \\
		\# unique words & 21,649 & 27,470& 24,546 &  22,099 \\
		\# OOV words  & 13,067 &  17,190& 15,724 & 14,331 \\ 	
		\# URLs  & 20,351 & 25,405 & 23,100 & 20,885   \\ 
		\# hashtags  & 6,784 & 8,019 & 7,869 & 7,346   \\ 
		\hline
	\end{tabular}
	\caption{Statistics of the TREC Microblog Track datasets.}
	\label{statistics}
\end{table}

\begin{table*}[t]
	\centering
	\footnotesize
		\begin{tabular}{| C{0.35cm} | c | l l | ll | ll | ll |}
			\hline
			\multirow{2}{*}{\textbf{ID}} &  \textbf{Model} & \multicolumn{2}{c|}{\textbf{2011}} & \multicolumn{2}{c|}{\textbf{2012}} & \multicolumn{2}{c|}{\textbf{2013}} & \multicolumn{2}{c|}{\textbf{2014}} \\
			& {\bf Metric} & {\bf MAP} & {\bf P30} & {\bf MAP} & {\bf P30} & {\bf MAP} & {\bf P30} &  {\bf MAP} & {\bf P30} \\
			\hline \hline
			\multicolumn{10}{|c|}{\textbf{Non-Neural Baselines}} \\
			\hline 
			1  & {QL} & 0.3576 &0.4000 & 0.2091 & 0.3311 & 0.2532 & 0.4450 & 0.3924 & 0.6182 \\
			2  & {RM3} & 0.3824$^1$ & 0.4211$^1$ & 0.2342$^1$ & 0.3452 &0.2766$^{1,2}$ & 0.4733$^{1}$\textit{} & \textbf{0.4480}$^{1,3}$ & 0.6339 \\ \hline
			3 & {L2R (all)} & 0.3845$^1$ & 0.4279 & 0.2291$^1$ & 0.3559 &0.2477 & 0.4617  & {0.3943} & 0.6200 \\
			&  {(text)} & 0.3547 & 0.4027 & 0.2072 & 0.3294 &0.2394 & 0.4456 & {0.3824} & 0.6091 \\
			&  {(text+URL)} & 0.3816 & 0.4272 & 0.2317 & 0.3667 &0.2489 & 0.4506 & {0.3974} & 0.6206 \\
			&  {(text+hashtag)} & 0.3473 & 0.4020 & 0.2039 & 0.3175 &0.2447 & 0.4533 & {0.3815} & 0.5939 \\
			\hline \hline
			\multicolumn{10}{|c|}{\textbf{Neural Baselines}} \\
			\hline 
			4 & {DSSM~\shortcite{huang2013learning}} & 0.1742 & 0.2340 & 0.1087 & 0.1791 & 0.1434 & 0.2772 & 0.2566 & 0.4261 \\
			5 & {C-DSSM~\shortcite{shen2014learning}} & 0.0887 & 0.1122 & 0.0803 & 0.1525 & 0.0892 & 0.1717 & 0.1884 & 0.2752 \\
			6 & {DUET~\shortcite{mitra2017learning} } &0.1533 & 0.2109 & 0.1325 & 0.2356 & 0.1380 & 0.2528 & 0.2680 & 0.4091 \\ 
			\hline
			7 & {DRMM~\shortcite{guo2016deep}} &0.2635 & 0.3095 & 0.1777 & 0.3169 &0.2102 & 0.4061  & 0.3440 & 0.5424 \\
			8 & {K-NRM~\shortcite{xiong2017end}} &0.2519 & 0.3034 & 0.1607 & 0.2966 &0.1750 & 0.3178 & 0.3472 & 0.5388  \\ 
			\hline
			9 & {PACRR~\shortcite{hui2017pacrr}} & 0.2856 & 0.3435 & 0.2053 & 0.3232 & 0.2627 & 0.4872 & 0.3667 & 0.5642 \\
			\hline \hline
			\multicolumn{10}{|c|}{\textbf{Neural Baselines with Interpolation}} \\
			\hline
			10 & {DUET+} & 0.3576 & 0.4000 & 0.2243$^{\scriptsize 1}$ & 0.3644$^{\scriptsize 1}$ & 0.2779$^{\scriptsize 1,3}$ & 0.4878$^1$ & 0.4219$^{\scriptsize 1,3}$ & \textbf{0.6467}$^{\scriptsize 1}$ \\  
			11 & {DRMM+} &0.3477 & 0.4034 &0.2213 & 0.3537 &0.2639 & 0.4772 &0.4042 & 0.6139 \\
			12 & {K-NRM+} &0.3576 & 0.4000 & 0.2277$^{\scriptsize 1}$ & 0.3520$^{\scriptsize 1}$ & 0.2721$^{\scriptsize 1,3}$ & 0.4756 & 0.4137$^{\scriptsize 1,3}$ & 0.6358$^{\scriptsize 1}$ \\
			13 & {PACRR+} &0.3810&0.4286$^{\scriptsize 1}$&0.2311$^{\scriptsize 1}$&0.3576$^{\scriptsize 1}$&0.2803$^{\scriptsize 1,3}$&0.4944$^{\scriptsize 1}$&0.4140$^{\scriptsize 1,3}$&0.6358$^{\scriptsize 1}$ \\
			\hline \hline
			\multicolumn{10}{|c|}{\textbf{Our Model}} \\
			\hline			
			14 & {MP-HCNN}& 0.3832 & 0.4075 & 0.2337$^{\scriptsize 1}$ & 0.3689$^{\scriptsize 1}$ & 0.2818$^{\scriptsize 1,3}$ & 0.5222$^{\scriptsize 1,3}$ & 0.4304$^{\scriptsize 1,3}$ & 0.6297 \\
			15 & {MP-HCNN+}& \textbf{0.4043}$^{\tiny 1,2,3}_{\tiny 12}$ & \textbf{0.4293}$^{\tiny 1}_{\tiny 12}$ & \textbf{0.2460}$^{\tiny 1,3}_{\tiny 12,13}$ & \textbf{0.3791}$^{\tiny 1,2,3}_{\tiny 12,13}$ & \textbf{0.2896}$^{\tiny 1,3}_{\tiny 12}$ & \textbf{0.5294}$^{\tiny 1,2,3}_{\tiny 12,13}$ & 0.4420$^{\tiny 1,3}_{\tiny 12,13}$ & 0.6394 \\
			 & & (+13.1\%) & (+7.3\%) & (+17.6\%) & (+14.5\%) & (+14.3\%) & (+18.9\%) & (+12.6\%) & (+3.4\%) \\
			\hline 
		\end{tabular}
		\caption{Main results on the TREC Microblog 2011--2014 datasets. Rows are numbered in the first column, where each represents a model or a contrastive condition. The last row shows the relative improvement against QL. The best numbers on each dataset are in bold. Superscripts and subscripts indicate the row indexes for which a metric difference is statistically significant at $p < 0.05$. Only methods 1--3 and 12--13 are compared with all other methods in the significance tests. }
	\label{tab:overall}
\end{table*}

\smallskip \noindent \textbf{Baselines.} We compare our model to a number of non-neural baselines as
well as recent neural ranking models designed for
``standard'' {\it ad hoc} retrieval tasks on web and newswire
documents (we call these the neural baselines).

The non-neural baselines include the most widely-used language model and pseudo-feedback methods:\
Query Likelihood (QL)~\cite{ponte1998language} and RM3~\cite{lavrenko2001relevance}.
We also compare to LambdaMART~\cite{burges2010ranknet}, the learning-to-rank model (L2R) that won the Yahoo!~Learning to Rank Challenge~\cite{burges2011learning}. We designed three sets of features:
(1) Text-based:\ In addition to QL, we compute another four overlap-based measures between each query-tweet pair:\ word overlap and IDF-weighted word overlap between all words and only non-stopwords, from Severyn and Moschitti~\shortcite{severyn2015learning};
(2) URL-based:\ whether the tweet contains URLs and the fraction of query terms that matched parts of URLs; 
(3) Hashtag-based:\ whether the tweet contains hashtags and the fraction of query terms that matched hashtags.

The neural baselines include recent state-of-the-art neural ranking models from the information retrieval literature. We compared to three sets of neural baselines: 
\begin{itemize}
\item Character-based:\ DSSM \cite{huang2013learning}, \mbox{C-DSSM} \cite{shen2014learning}, DUET \cite{mitra2017learning} 
\item Word-based:\ DRMM \cite{guo2016deep}, K-NRM \cite{xiong2017end} 
\item Word ngram-based:\ PACRR \cite{hui2017pacrr}
\end{itemize}

\smallskip \noindent \textbf{Implementation Details.} 
We apply the same padding strategy to the four datasets based on the
longest (query, tweet) length in the datasets.
The URLs are truncated and padded to 120 characters.
Mentions are removed and hashtags are treated as normal words (i.e., ``\#bbc'' to ``bbc'').
The IDF weights of word and character $k$-grams are computed from 
the Tweets2013 collection~\cite{lin2013overview}, which consists of 243 million tweets crawled from 
Twitter's public sample stream between February~1 and March~31, 2013.

To enable fair comparisons with the baselines, we adopt the same training
strategies in all our experiments, including embeddings, optimizer, and hyper-parameter settings.
We used trainable word2vec embeddings~\cite{mikolov2013distributed} with a learning rate of 0.05
and the SGD optimizer. We randomly initialize the embedding of OOV words and
character trigrams between $[0, 0.1]$.
The number of convolutional layers $N$ is set to $4$. 
We tune the number of convolutional filters and batch size in $[256, 128, 64]$ and the dropout rate between $0.1$ and $0.5$.
The interpolation parameter $\lambda$ (with the QL score) is tuned after the neural network model converges. 
Our code and data are publicly available,\footnote{https://github.com/jinfengr/neural-tweet-search} while other neural baselines can be found in the MatchZoo library.\footnote{https://github.com/faneshion/MatchZoo}


\begin{table*}[t]
	\centering
	\footnotesize
	\begin{tabular}{| l | ll | ll | ll | ll |}
		\hline
		\textbf{Setting} & \multicolumn{2}{c|}{\textbf{2011}} & \multicolumn{2}{c|}{\textbf{2012}} & \multicolumn{2}{c|}{\textbf{2013}} & \multicolumn{2}{c|}{\textbf{2014}} \\
		{\bf Metric} & {\bf MAP} & {\bf P30} & {\bf MAP} & {\bf P30} & {\bf MAP} & {\bf P30} &  {\bf MAP} & {\bf P30} \\
		\hline \hline		
		{MP-HCNN} & \bf 0.3832 & \bf0.4075 & \bf 0.2337& \bf0.3689 & \bf 0.2818 & \bf 0.5222 & \bf 0.4304 &\bf 0.6297 \\ 
		\hline \hline
		{$-$ mean pooling} & 0.3687$\star$ & 0.4054 & 0.2251 & 0.3480 &0.2766 & 0.5000 &0.3907$\star$ & 0.5897$\star$ \\ 
		{$-$ max pooling} & 0.0982$\star$ & 0.1320$\star$ & 0.0767$\star$ & 0.1243$\star$ & 0.0920$\star$ & 0.1706$\star$ &0.1934$\star$& 0.2176$\star$ \\
		{$-$ IDF weighting} & 0.3511$\star$ & 0.3714$\star$ &0.2119$\star$ &0.3452 & 0.2717$\star$ & 0.4967$\star$ & 0.3992 & 0.6097$\star$\\
		{$-$ word module} & 0.1651$\star$ & 0.1293$\star$ &0.0762$\star$ & 0.1119$\star$ & 0.0987$\star$ & 0.1517$\star$ & 0.1849$\star$ & 0.2048$\star$ \\
		\hline
		{$-$ URL char rep.} & 0.3594$\star$ & 0.3707$\star$ & 0.2131$\star$ & 0.3333$\star$ & 0.2797$\star$& 0.4989$\star$& 	0.4037$\star$ &0.6085$\star$ \\
		{$-$ doc char rep.} &0.3603$\star$& 0.3721$\star$& 0.2188$\star$ & 0.3537$\star$ & 0.2757$\star$ & 0.5122 &0.4012 & 0.6103\\
		{$-$ all char rep.} &0.3528$\star$& 0.3709$\star$ & 0.2087$\star$ &0.3271$\star$ &0.2718$\star$ &0.5011$\star$ & 0.4050$\star$ & 0.6091$\star$ \\
		\hline
	\end{tabular}
	\caption{Ablation Study. $\star$ denotes scores significantly lower than the MP-HCNN model at $p < 0.05$.}
	\label{tab:ablation}
\end{table*}

\section{Results}

Our main results are shown in Table \ref{tab:overall}. 
Rows are numbered in the first column, where each represents a model or a contrastive condition.
We compare our model to three sets of baselines: non-neural, neural, and interpolation.
Interpolation methods are denoted by a symbol ``+'' at the end of the original model name, such as DUET+.
We run statistical significance tests using Fisher's two-sided, paired randomization test~\cite{smucker2007comparison} against the three non-neural baselines:\ QL, RM3, and L2R (with all features), and the best neural baselines:\ K-NRM+ and PACRR+.
Superscripts and subscripts indicate the row indexes for which a metric difference is statistically significant at $p < 0.05$. 

From the first block ``Non-Neural Baselines'' in Table~\ref{tab:overall},
we can see that RM3 significantly outperforms QL on all
datasets, demonstrating its superior effectiveness. However, RM3 requires
an additional round of retrieval to select terms for query expansion, and thus is
substantially slower. Lambda\-MART achieves effectiveness on par with RM3
when using all the hand-crafted features. From its contrastive variant with only
text-based features, we can see that the overlap-based features provide little gain over QL. 
Comparing the rows ``(text+URL)'' and ``(text+hashtag)'' to row ``(text)'', adding URL-based features
leads to a significant improvement over text-based features, while hashtag-based features
seem to bring fewer benefits. This confirms our observation (cf.~Table~\ref{statistics}) that
URLs appear frequently in tweets and contain meaningful relevance signals.

Looking at the second block ``Neural Baselines'', 
we find that all the neural methods perform worse than the QL baseline, 
showing that existing neural ranking models fail to adapt to tweet search. 
In fact, all the character-based approaches (DSSM, C-DSSM, DUET) are consistently 
worse than the word-based approaches 
(DRMM, K-NRM). This is likely attributable to the fact that all word-based NN models
use pre-trained word vectors that encode more semantics than a random initialization 
of character trigram embeddings, suggesting that the Twitter datasets are not sufficient to
support learning character-based representations from scratch. 
Particularly, C-DSSM suffers 
more than DSSM, showing that a more complex model leads to lower effectiveness in a 
data-poor setting. 

Comparing word-based models (DRMM, K-NRM) with ngram-based models (PACRR), 
we see that PACRR performs much better by modeling ngram semantics. 
In addition, the small parameter space of DRMM (1541 parameters in total) 
suggests that the low effectiveness of neural baselines
is not simply due to a shortage of data. In comparison, our MP-HCNN achieves
high effectiveness on all datasets for both metrics, beating all neural baselines by a large margin. 
We believe that effectiveness gains mainly come from two aspects:\ 1) Unlike all neural baselines that model the similarity matrix computed from the product of query and document embeddings, our approach directly models the raw texts and better preserves semantic representations after hierarchical convolutional operations; 2) Character-level modeling provides additional relevance signals.

In the third block ``Interpolation Baselines'', we observe that simple interpolation with QL
boosts the effectiveness of all neural baselines dramatically, showing that exact match signals
are complementary to the soft match signals captured by NN methods. This observation also holds
for our MP-HCNN+, although the margin of improvement is smaller due to the effectiveness of MP-HCNN alone.
The best results on the TREC Microblog 2011--2013 datasets are
obtained by MP-HCNN+, with an average of 14.3\% relative improvement against QL (shown in the last row). 
Also, MP-HCNN+ is significantly better than all the best baselines in most settings, 
except for TREC 2014, where the QL baseline already achieves fairly high effectiveness in absolute terms, limiting the space for potential improvement. 

\subsection{Ablation Study}

To better understand the contribution of each module in our proposed model, we perform an
ablation study on the base MP-HCNN model, removing each component step by step. 
Here, we aim to study how the semantic-level, character-level, and weighting modules
contribute to model effectiveness. 
Results on the TREC 2011--2014 datasets are shown in Table~\ref{tab:ablation}, with each row denoting the removal of
a specific module. For example, the row ``$-$ URL char rep.''\ represents removing
the URL modeling module. 
The $\star$ symbol denotes that the model's effectiveness in an ablation setting is significantly 
lower than MP-HCNN at $p < 0.05$. 

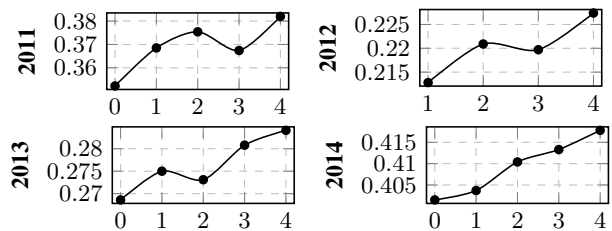
\begin{figure}
\centering
\begin{subfigure}{0.22\textwidth}
	\begin{tikzpicture}
	\begin{axis}[width=4cm, height=2.6cm,line width=0.6,enlargelimits=0.05,ylabel=2011,ytick scale label code/.code={$\times 10^{#1}$}, /pgf/number format/precision=4
	,label style={font=\bfseries\small},tick label style={font=\bfseries\small},grid=major, grid style=dashed]
	\addplot[smooth,mark=*,mark size=1.5pt,color=black] coordinates {(0, 0.3522)
 (1, 0.3685)
 (2, 0.3754)
 (3, 0.3674)
 (4, 0.3820)};
	\end{axis}
	\end{tikzpicture}
\end{subfigure} 
\begin{subfigure}{0.22\textwidth}
	\begin{tikzpicture}
	\begin{axis}[width=4cm, height=2.6cm,line width=0.6,enlargelimits=0.05,ylabel=2012,ytick scale label code/.code={$\times 10^{#1}$}, /pgf/number format/precision=3
	,label style={font=\bfseries\small},tick label style={font=\bfseries\small},grid=major, grid style=dashed]
	\addplot[smooth,mark=*,mark size=1.5pt,color=black] coordinates {(1, 0.2128)
(2, 0.2209)
(3, 0.2197)
(4, 0.2274)};
	\end{axis}
	\end{tikzpicture}  
\end{subfigure}

\begin{subfigure}[b]{0.23\textwidth}

	\begin{tikzpicture}
	\begin{axis}[width=4cm, height=2.6cm,line width=0.6,enlargelimits=0.05,ylabel=2013,ytick scale label code/.code={$\times 10^{#1}$}, /pgf/number format/precision=3
	,label style={font=\bfseries\small},tick label style={font=\bfseries\small},grid=major, grid style=dashed]
	\addplot[smooth,mark=*,mark size=1.5pt,color=black] coordinates {(0, 0.2686)
(1, 0.2750)
(2, 0.2731)
(3, 0.2808)
(4, 0.2841)};
	\end{axis}
	\end{tikzpicture}
\end{subfigure}
\begin{subfigure}[b]{0.22\textwidth}

	\begin{tikzpicture}
	\begin{axis}[width=4cm, height=2.6cm,line width=0.6,enlargelimits=0.05,ylabel=2014,ytick scale label code/.code={$\times 10^{#1}$}, /pgf/number format/precision=3
	,label style={font=\bfseries\small},tick label style={font=\bfseries\small},grid=major, grid style=dashed]
	\addplot[smooth,mark=*,mark size=1.5pt,color=black] coordinates {(0, 0.4015)
(1, 0.4037)
(2, 0.4104)
(3, 0.4133)
(4, 0.4178)};
	\end{axis}
	\end{tikzpicture}  
\end{subfigure}
\caption{MAP with different convolutional depth $N$ on TREC 2011--2014 datasets.}
\label{cnn-depth}
\end{figure}

\begin{figure*}[t]
	\begin{subfigure}{0.45\textwidth}
		\includegraphics[width=2.2cm, height=8.6cm,angle=-90]{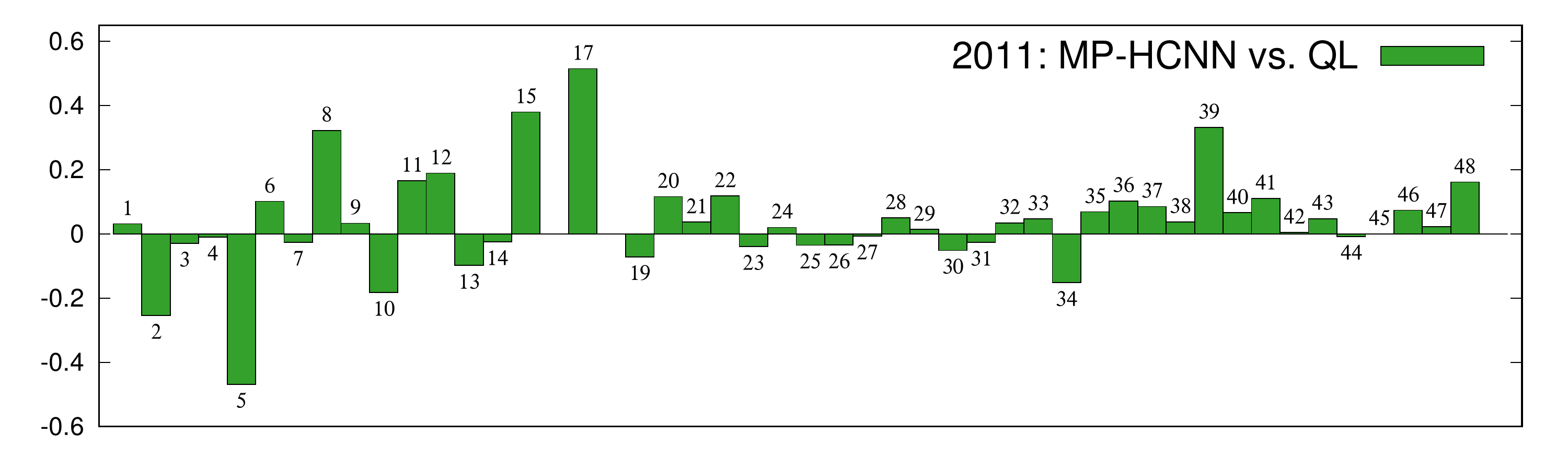}
	\end{subfigure} \qquad
	\begin{subfigure}{0.45\textwidth}
		\includegraphics[width=2.2cm, height=8.6cm,angle=-90]{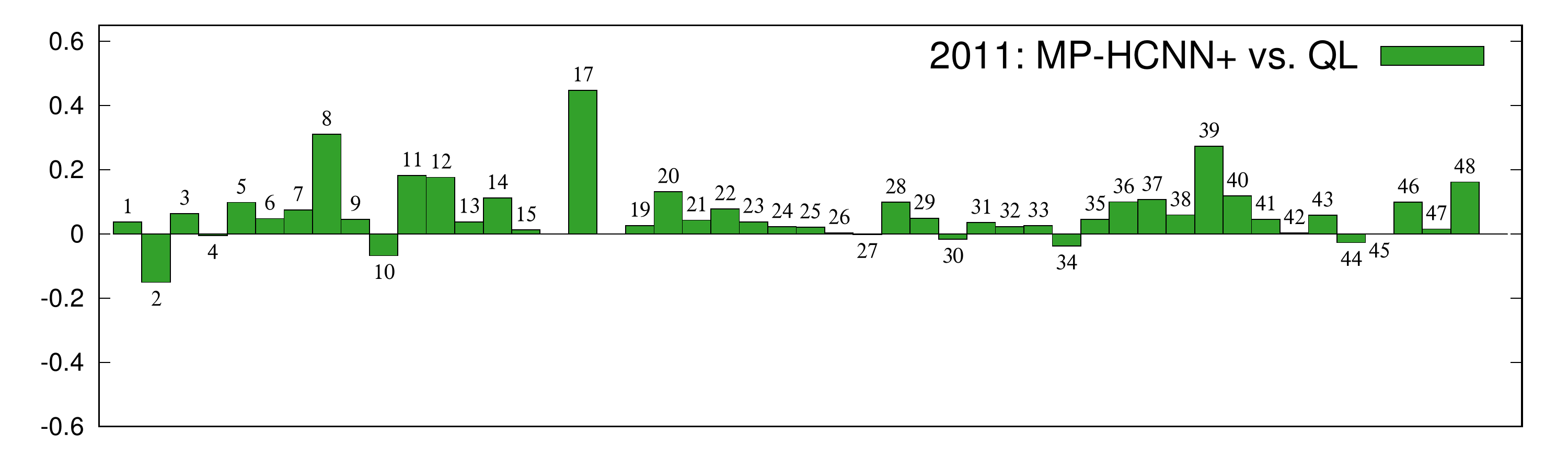}
	\end{subfigure}
	\caption{Per-topic MAP differences of MP-HCNN and MP-HCNN+ vs.\ QL on TREC 2011.}
	\label{per-topic}
\end{figure*}

\begin{table*}[ht]
	\scriptsize
	\centering
	\begin{tabular}{|c|C{10.5cm}|c|c|c|}
		\hline
		\multirow{2}{*}{\textbf{ID}} & \multirow{2}{*}{\textbf{Sample Tweet}} & \multirow{2}{*}{\textbf{Label}} & \multicolumn{2}{c|}{\textbf{Score(Rank)}} \\ \cline{4-5} 
		& &  & \textbf{QL} & \textbf{MP-HCNN} \\ \hline
		    1 & \#ps3 best sellers: fifa soccer 11 ps3 \#cheaptweet https://www.amazon.com/fifa-soccer-11-playstation-3 & I & 7.33(\#54) & 0.85(\#1) \\ \cline{1-1} \cline{2-5} 
    2 & qatar 's 2022 fifa world cup stadiums: 
https://wordlesstech.com/qatars-2022-fifa-world-cup-stadiums/ & R & 10.58(\#2) & 0.41(\#105) \\ \cline{1-1} \cline{2-5} 
    3 &  2022 world cup could be held at end of year:  fifa : lausanne switzerland  the 2022 world cup in qatar: http://www.reuters.com/article/us-soccer-world-blatter & R & 11.25(\#1) & 0.31(\#127) \\ \hline
	\end{tabular}
	\caption{Sample analysis of the worst-performing topic~2 (``2022 fifa soccer''). \textbf{I} denotes irrelevant and \textbf{R} denotes relevant.}
	\label{samples}
\end{table*}

From the first two rows ``$-$ mean/max pooling'', we can see that removing max pooling leads to a 
significant effectiveness drop while removing mean pooling only results in a minor reduction.
Also, removing the IDF weights makes the results consistently and significantly worse
across all four datasets, which confirms that injecting external weights is important for tweet search. 
It is also no surprise that
the complete word-level module is essential to capture relevance, as shown in the table.

Turning our attention to the last three rows, we observe that removing the character representations 
of URLs or documents both lead to significant drops across all datasets, with larger drops
when URLs are removed. This suggests that URLs
provide more relevance signals than character-level document modeling. 
Taking away the entire character-level module causes slightly more effectiveness loss.
To conclude, the word-level matching module contributes the most effectiveness, 
but the character-level matching module still provides complementary and significantly useful signals.
However, given the low effectiveness of the character-based models in Table~\ref{tab:overall}, we add
a caveat:\ with more training data or pre-trained character embeddings,
we would expect the benefits of the character-level matching module to increase.

We also examine how the depth of the hierarchical convolutional layers affects model effectiveness.
Figure~\ref{cnn-depth} shows effectiveness in terms of MAP with different convolutional depth $N$ 
on the TREC 2011--2014 datasets. 
A setting of $N=0$ means that there are no convolutional layers on top of the
embedding layer, and the prediction is purely based on matching evidence at the word-level. A larger
value of $N$ indicates that longer phrases are captured and represented. We
can clearly see that there is a consistent gain in effectiveness with increasing depths on the datasets,
except for $N=3$ on TREC 2011. Here, the improvement
at $N = 2$ is already quite close to the upper bound at $N=4$. This suggests that modeling
short phrases brings immediate benefit while the inclusion of longer phrases 
only marginally boosts overall effectiveness.
In summary, this ablation experiment clearly shows the value of our hierarchical design in semantic modeling at the phrase level.

\subsection{Error Analysis}

So far, we have shown that our weighted similarity measurement component, as well as the 
URL matching and phrase matching components (enabled by the hierarchical architecture),
are crucial to our model's effectiveness. However, we still lack knowledge about
the following two questions:\ (1) What are the common characteristics of
well-performing topics, and how do the different components 
contribute to overall effectiveness? (2) When does our model fail, and how
can we further improve the model? To answer these questions, we provide additional qualitative and 
quantitative analyses over sample tweets from well-performing and poor-performing topics. 

In Figure~\ref{per-topic}, we visualize per-topic differences in terms of MAP for
MP-HCNN and MP-HCNN+ against the QL baseline on the TREC 2011 dataset. Since other datasets exhibit similar trends, we omit their figures here. Overall, we see that the
MP-HCNN model shows improvements for the majority of topics.
In total, MP-HCNN wins on 26 topics and loses on 13 topics out of 49 topics. 
The average margin of improvement is also greater than the losses.
With the interpolation technique, MP-HCNN+ is able to smooth out
the errors in many poor-performing topics, such as topic 5 ``nist computer security'', resulting in more
stable improvements.  

In addition, we select the five best-performing topics (15, 17, 39, 91, 105) from the TREC 2011 
and 2012 datasets. For each topic, we select the top 20 tweets 
with the highest MP-HCNN prediction scores for analysis. 
We manually classify the matching evidence of the 100 selected tweets 
into the following categories (a tweet can satisfy multiple categories):
1) exact word match; 2) exact phrase match; 3) partial paraphrase match and
4) partial URL match, where partial match means that part of the tweet or URL
matches query terms.

%
%
%
%
%

\begin{table}
\begin{center}
	\begin{tabular}{lr}
		\hline
		Category & Percentage (\%) \\
		\hline
		Exact word match   & 100 \\
		Exact phrase match & 44 \\ 
		Partial paraphrase match & 59 \\
		Partial URL match &  29 \\
		\hline
	\end{tabular}
\end{center}
	\caption{Matching evidence breakdown by category based on manual analysis of the top 100 tweets for the five best-performing topics.}
	\label{matching-evidence}
\end{table}

Table~\ref{matching-evidence} provides a breakdown of matching evidence by category. We can see that all tweets have exact word matches to the queries, 
and partial paraphrase matches occur more frequently than exact phrase matches, 
suggesting that our hierarchical architecture with embedding inputs
is able to capture those soft semantic match signals. 
In addition, partial URL matches make up another
big portion, affirming the need for character-level URL modeling.

To gain additional insights into how our model fails, 
we analyze some sample tweets for the worst-performing topic~2
(``2022 fifa soccer''), shown in Table~\ref{samples}. Column ``Label'' represents whether the tweet
is relevant to the query:\ ``R'' denotes relevant and ``I'' denotes irrelevant. Column
``Score(Rank)'' shows the prediction scores and the rank position of sample tweets 
by each method (QL or MP-HCNN). 


Looking at the first tweet, it obtains the highest score by MP-HCNN due to the phrase 
match ``fifa soccer'' (a match score of 0.89 from the softmax at the similarity measurement layers) for the content and URL. However, the MP-HCNN fails to
understand that ``fifa soccer 11'' refers to a video game on the PS3, showing the limits of 
a matching-based algorithm for entity disambiguation. 
In contrast, though the second and third tweets look more relevant to the query, 
they are assigned much lower scores by the MP-HCNN. This is because 
the query term ``2022'' is an out-of-vocabulary word, and thus its matching evidence is 
greatly reduced due to the random initializations of OOV word embeddings.
The semantic match of the phrase ``world cup'' to the query has a low match score of 0.36, which doesn't help boost its overall relevance. 

In summary, results from these manual analyses confirm the
quantitative results from the previous sections. Exact term match
remains critical to relevance modeling, while soft matches that
incorporate phrases and semantic similarities make substantial
contributions as well. Furthermore, although URLs play a smaller
role in matching, they provide complementary signals.
Though soft-match signals can be led astray, as our error
analysis shows, overall they help more than they hurt.

\section{Conclusions}

To conclude, this paper presents, to our knowledge, the first substantial work on
neural ranking models for {\it ad hoc} retrieval on social media.  We
have identified three main characteristics of social media posts that
make our problem different from ``standard'' document ranking over web
pages and newswire articles. Our model is specifically designed to cope
with each of these issues, capturing multiple signals from queries,
social media posts, as well as URLs contained in the posts, at the
character-, word-, and phrase-levels. Extensive experiments
demonstrate the effectiveness of our model and ablation studies
verify the importance of each model component, suggesting that our
customized architecture indeed captures the characteristics of our
domain-specific ranking challenge.

\bibliography{AAAI-RaoJ.6198}
\bibliographystyle{aaai} 

\end{document}